# Mixed-resolution hybrid modeling in an element-based framework

Kara Bocan and Natasa Miskov-Zivanov
University of Pittsburgh
Pittsburgh, PA, USA, 15213
Email: knb12@pitt.edu

*Abstract*—Computational modeling of a complex system is limited by the parts of the system with the least information. While detailed models and high-resolution data may be available for parts of a system, abstract relationships are often necessary to connect the parts and model the full system. For example, modeling food security necessitates the interaction of climate and socioeconomic factors, with models of system components existing at different levels of information in terms of granularity and resolution. Connecting these models is an ongoing challenge. In this work, we demonstrate methodology to quantize and integrate information from data and detailed component models alongside abstract relationships in a hybrid element-based modeling and simulation framework. In a case study of modeling food security, we apply quantization methods to generate (1) time-series model input from climate data and (2) a discrete representation of a component model (a statistical emulator of crop yield), which we then incorporate as an update rule in the hybrid element-based model, bridging differences in model granularity and resolution. Simulation of the hybrid element-based model recapitulated the trends of the original emulator, supporting the use of this methodology to integrate data and information from component models to simulate complex systems.

## I. Introduction

THERE are increasing challenges associated with global food security and the exacerbating effects of climate change [1], [2], [3]. According to the World Health Organization, "More than 820 million people in the world are hungry today, underscoring the immense challenge of achieving the Zero Hunger target by 2030" [3]. Food security is a complex issue spanning multiple domains, where detailed models are available for some factors but not for others. While there exist very detailed models of climate and crop yield, meaningful analysis of food security requires inclusion of socioeconomic factors and less quantifiable concepts such as safety and inequality [4], [2], [3].

This modeling landscape is not unique to food security. Models exist at various levels of *granularity* – complexity of the components of the model – and *resolution* – precision of the values of components in the model [4], [1]. At the most detailed, *process-based* or *mechanistic models* are based on physical laws governing behavior in a system [4], [5], [6]. However, it is often necessary to work at higher levels of abstraction (lower granularity and/or resolution) for computational practicality, due to lack of availability of process-based model results, or due to a lack of information required to develop process-based models [4], [6], [7], [8]. A complete model of a complex system is ultimately limited by the parts of the system with the least information.

Abstract representations can bridge these knowledge inconsistencies, while also aiding interpretation of models and simulation results even where detailed models are available [5], [7], [9]. Abstraction can reduce computational requirements of simulation, enabling inclusion of additional system components and rapid simulation with varying inputs and parameters. One such abstract formalism is *element-based modeling*, where a model consists of multiple *elements* with defined interactions, discrete variables are associated with each element, and *update rules* allow simulation of the model by calculating next-state values of each element variable over time [10], [11], [12]. The definition of model elements determines granularity, and the number of discrete levels for each variable determines resolution, enabling flexibility in representation. However, there are challenges associated with abstraction, and especially in linking abstract representations with detailed models [8].

*Hybrid models* have addressed some of the challenges of combining different model representations or modeling formalisms [13], [14], [15]. The term "hybrid modeling" has been used generally to describe any combination of different modeling approaches or formalisms [13], [14], and more formally to refer to a combination of process-based and data-driven models (distinguished as *hybrid semi-parametric modeling*) [15], [16]. Stephanou and Volpert in a review of hybrid modeling define three classes of hybrid models based on the connectivity and formalisms involved: (1) *decoupled* approaches comparing different models in parallel; (2) *coupled* models that are connected by specific variables, often used in multi-scale modeling applications; and (3) *intricate* models constructed such that one or more subcomponents are fully integrated within another formalism [13]. While hybrid models have been developed that connect or incorporate models of different formalisms, there is still a need for a general methodology to adapt and integrate information from multiple detailed models within an abstract modeling formalism.

In this work, we present methodology to combine information from detailed models and abstract relationships in a *hybrid element-based model*. Our hybrid modeling approach falls under the *intricate* class of hybrid models, integrating *component models* within the formalism of element-based modeling to achieve *mixed-resolution modeling*. In this approach we apply quantization methods to generate discrete-value abstract representations of component models suitable for connection to other elements and simulation in the element-based framework. These component models can be process-based or data-driven.

In this paper, we first describe our methodology for quantizing and integrating component model representations to form hybrid element-based models. Then, we present a case study demonstrating our methods applied to a model of food security and exploring the effects of mixed resolution of interacting elements in the model. Component models in this context are expected to be detailed models of processes such as climate and crop yield, and our methodology enables integration of these models with more abstract relationships representing factors such as conflict and displacement.

## II. BACKGROUND

### A. Element-Based Modeling

In an element-based modeling framework, summarized in **Fig. 1**, a model consists of multiple *elements*, analogous to nodes in a graphical representation, where each element represents a component of the system or is an abstract representation of aggregated system components. Elements in a model are connected by influences, analogous to directed edges in a graphical representation, indicating an *interaction* among these elements.

Each element has an associated value represented by a discrete variable. This value quantifies some aspect of the element, such as amount or activity. The value of each variable can take one of a defined number of discrete levels. In this way, element-based models are an extension of Boolean models, where variables can take two values representing ON/OFF or LOW/HIGH [7], [17]. Previous work has extended Boolean models to allow variables with three levels representing OFF/LOW/HIGH or LOW/MEDIUM/HIGH [12], [18], [19], [20]. This formalism uses abstract quantitative value representation, in contrast to values indicating truth or probability [21], [22].

The values of all element variables in a model at a given time represent the model *state*. In an element-based model, the next-state value of a given element variable is defined by its *update rule*, which is a function of the current-state values of all elements directly influencing the target element and the target element's current-state value. Update rules vary in complexity depending on how much information is available about the combined effects of influences on the target element. Thus, the element-based framework is designed to handle lack of information about parts of a system.

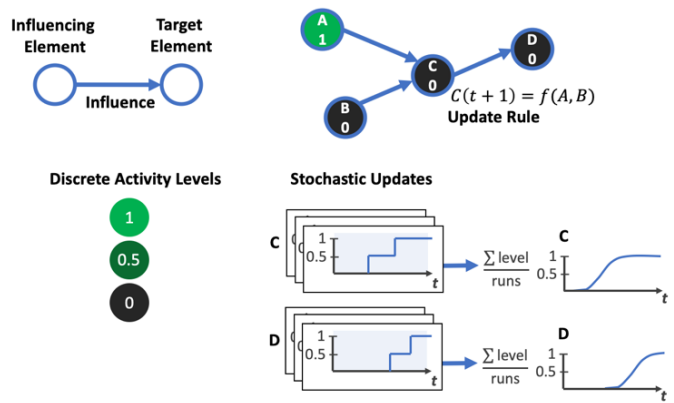

**Fig. 1.** Element-based modeling and simulation.

Simulation of element-based models is performed by choosing one or more elements to be updated at each simulation time step and updating the values of the chosen elements' variables according to their update rules [12]. In addition to update rules, an initial value is defined for each element for a given simulation, and the set of initial values for all elements for a given simulation defines a *scenario*. Elements updated at each time step can be chosen randomly for stochastic simulation. Multiple stochastic simulation runs are performed in order to inspect average behavior of elements in the model over time.

### B. Component Models

The goal of this work is to use component models to inform element-based model update rules, in combination with less complex update rules where component models are not available. There are likely many component models compatible with this methodology. Here we focus on *emulators* as abstractions of computationally intensive (e.g., process-based) models.

In the most basic form, emulators may consist of cached results of a computationally intensive model, providing a mapping of input to output without the need to run the process-based model. However, it may not be practical to run all necessary input combinations and generate output to create a complete cache of results. Statistical emulators can be trained on the results of process-based models to recapitulate the process-based model output as well as approximate missing values, avoiding the need to run the process-based model for all input combinations [4], [5], [6].

In addition to being less computationally intensive, emulators can provide abstraction by determining whether effects of some factors can be approximated or are negligible for a given context. For example, [5], [6] focused on effects of temperature and precipitation in developing emulators modeling impacts of climate change on crop yields.

## C. Quantization

As previously described, the element-based modeling framework represents element variable values in terms of discrete levels to enable lightweight simulation. If a detailed relationship such as an emulator is to be adapted for this framework, the values must therefore be quantized. Quantization methods enable conversion from either continuous values or high-resolution value ranges into a smaller number of discrete levels. More detail on our quantization methodology specific to this work is provided in the Methods.

## III. METHODS

Integrating component models into a hybrid element-based modeling framework necessitates vertically traversing levels of abstraction, such that component models are "pulled up" in abstraction to integrate with the element-based model. This traversal is necessary to modeling and simulating where the hybrid model includes parts of the system with varying detail in knowledge and information. Emulators are already an abstraction of process-based models, so in this work we are effectively abstracting further to the level of hybrid element-based models.

In our methodology, summarized in **Fig. 2** we apply quantization methods to incorporate component models into hybrid element-based models. Connecting component models requires resolving differences in model granularity and resolution. In the following sections, we first describe functionality added to the element-based modeling framework to support integration of component models. We then discuss our general approach to address the challenges of differing granularity and resolution, before presenting a case study demonstrating an application of our methodology to a model of food security.

### A. Hybrid Element-based Modeling

In this section, we describe functionality added to the element-based modeling framework enabling integration of component models to create a hybrid element-based model. In the element-based framework, we introduce: (1) setting of element values according to a time series, (2) lookup tables as update rules, and (3) varying resolution of discrete variables.

1) **Time-Series Set Values**

For simulation of element-based models, initial values are defined for each element variable, and at each time step in the simulation a number of element variables are selected to be updated according to their update rules. In addition to a single initial value, the value of an element variable can be set at any time step in the simulation, forcing the variable to take the designated value at the designated time step. Building on the functionality demonstrated in [12], these values can be defined in the form of a time series, such that a sequence of values is defined at corresponding simulation time steps.

In the element-based framework, we define model *inputs* as elements that do not have any upstream influences. A time series can be defined for a model input to set the values of that element variable over the course of the simulation. If a time series is defined for an element that is not an input, then the value of the variable will be forced to the set values at time steps included the time series, while the value at other time steps will be determined according to the influences on the element and its update rule.

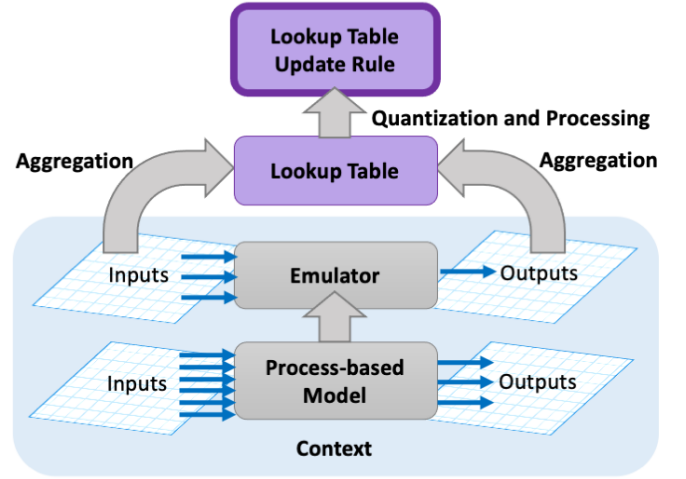

**Fig. 2.** Methodology to generate update rules from detailed models.

2) **Lookup Table Update Rules**

Prior to this work, all update rules in the element-based framework were *incremental update rules*, calculated as follows:

- A *positive score* is calculated according to a defined function of the positive influences, and a negative score is calculated according to a defined function of the negative influences.
- The element's next-state value is determined from the current-state value and comparison of the positive and negative scores: the value is incremented by one level if the positive score is greater than the negative score, or the value is decremented by one level if the positive score is less than or equal to the negative score.

The functions of positive or negative influences are defined using notation that can specify discrete AND/OR (similar to min/max fuzzy logics [21], but in a discrete value framework), nested interactions (an element can influence an interaction between other elements), and weighted summation [11], [12].

There are special cases for these incremental update rules when an element lacks either positive or negative influences. If an element lacks both positive and negative influences, it has no upstream elements, and therefore its value is only updated according to user input in the form of an initial value or time series. If an element has positive influences but lacks negative influences, when the positive score is zero, although the score comparison will be equal (because the negative score will always be zero), the value is decreased to compensate for the lack of negative influences. Similarly, if an element has negative influences but lacks positive influences, when the negative score is zero, the value is increased.

The current work preserves this incremental update rule notation, while adding functionality to support lookup tables as update rules to integrate quantized input/output mappings representing component models. To represent a valid update rule in the element-based framework, the lookup table must include every possible combination of discrete input values and the associated discrete output. This is an extension of truth tables describing logic functions to represent functions of discrete variables.

While these lookup table update rules can incorporate information from detailed models, incremental update rules are still valuable where this amount of information is not available. Therefore, in the same element-based model, some elements may have lookup tables as update rules, while other elements may have incremental update rules; this further allows update rules to be defined according to the amount of information available about how influences combine in affecting a target element.

3) **Discrete Variable Resolution**

In the hybrid element-based model, we allow variables with mixed resolution. That is, each element's variable may have a different number of discrete levels. The number of levels is fixed for a given element variable.

To avoid a variable with more levels inherently having greater weight in an update rule, the value of each variable is normalized to between 0 and 1. The number of discrete levels therefore defines the resolution of the variable, and integer weights are used to control relative strengths of influences in update rules. From a quantization perspective, each discrete level can be thought of as a bin, representing a range of real number values before quantization.

Additionally, to address the effects of mixed resolution variables on incremental update rules, we calculate the increment proportionally to the number of levels for an element and the difference between positive and negative scores.

With the aforementioned functionality added to the element-based framework, we next address how to connect component model inputs and outputs to other elements in the model, which involves mapping across different levels of granularity and resolution.

### B. Granularity

Even at a similar level of abstraction, it is a challenge to connect variables among different detailed models [23]. It is an additional challenge to align variables while vertically traversing levels of abstraction. The mapping of variables across levels of abstraction is likely not one-to-one, and requires concept matching, aggregation, and potentially resolving mismatched functions or connections among model variables.

In this work, we utilize the abstraction provided by emulators and spatial aggregation to map the component model inputs and outputs to elements in the element-based model representation, but additional work is needed to fully address differences in granularity.

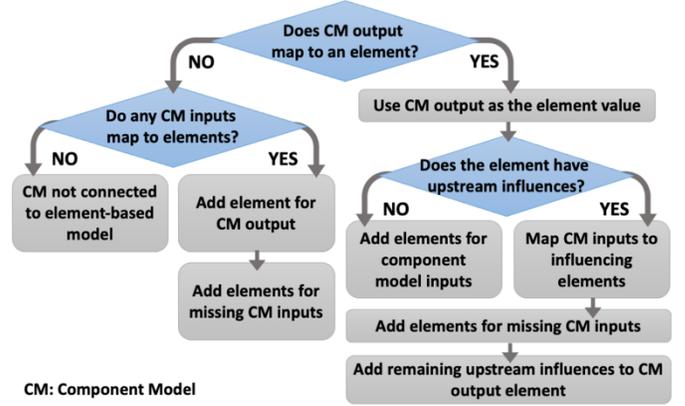

**Fig. 3.** Mapping of component model variables to element-based models.

The likely case is that some component model variables correspond to higher-abstraction-level elements, but the full set of component model inputs/outputs does not directly correspond to element influences. Our proposed strategy is summarized in **Fig. 3**.

### C. Resolution and Quantization

To adapt a component model for the element-based framework, we apply quantization methods to represent the inputs and outputs of the component model with discrete values, effectively converting each variable in the component model into a discrete variable in the element-based model. Quantization thresholds determine the range of values represented by the discrete variable as well as the variable's resolution (number of discrete levels). Quantization can be uniform or non-uniform; here we focus on uniform quantization.

To quantize each component model variable, we need to determine quantization thresholds. We first inspect the range of possible values of component model variables to define minimum and maximum quantization thresholds. Then, we choose intermediate thresholds to determine the discrete variable resolution. While in this work we choose a number of levels and define uniform quantization thresholds, the choice of quantization thresholds can be further informed by context (e.g., high or low rainfall could depend on region, season, type of crop), causal effects (e.g., does the effect of an input asymptote outside a certain range), or availability of results (e.g., limitations of an emulator to the range of process-based model results). It will likely make sense to have different resolution for different variables. As we will show in the case study, the effects of using higher resolution for component model variables in the hybrid element-based model will ultimately depend on the resolution of other element variables in the model.

The goal of this quantization methodology is to produce a lookup table representing the component model input/output in terms of discrete levels, to be used as an update rule in the hybrid element-based model. This lookup table must consist of output values associated with each possible combination of

45input values, such that there is a next-state value defined for all values of the variables in the element's update rule. We therefore use quantization methods to generate a finite list of component model input combinations associated with output. Emulators are useful in this approach because the output for a given combination of inputs is relatively simple to generate. However, there is also the choice of whether and how to aggregate output across the range of input values represented by a discrete level, or to choose representative input values (e.g., the middle of the value range) to generate output with an emulator.

Other potential issues relate to differences in spatial and temporal resolution of component models and element-based models:

1) **Spatial Resolution**

The element-based framework represents spatial or location information with abstraction, where one element can represent an aggregation of values for a given area. For example, rather than using multiple elements to represent crop yields for every harvest area in a given region individually, we can use just one element representing an aggregation of the crop yield across all harvest areas in the region (e.g., average crop yield per hectare). With this framework, we can create multiple elements representing locations at the desired level of spatial resolution. There also exist spatial aggregation tools for component models [24], [25]. However, these tools first calculate values at high spatial resolution before performing aggregation; that is, the component model inputs are at higher spatial resolution than the aggregated output. This precludes use of element-based model variables directly as component model inputs when the spatial resolution is not sufficient to determine location-based variable values (e.g., grid cell effects required to determine crop yields in [24]). To address this issue in our case study, we spatially aggregate the component model inputs to further abstract the component model for integration with the element-based model.

2) **Temporal Resolution**

Models may use temporal averages as input and produce outputs at different temporal resolution. For example, emulators in [5] use monthly averages as input, and produce annual estimates of crop yields. The abstract representation of values in the element-based framework allows element values to represent amount in terms of a temporal average. However, this temporal resolution should be consistent for connections of that element to other elements in the model, and with the time scale of the simulation. In the element-based framework, delays or update rates can be adjusted to account for elements with different time scales [12], [26]. Another option is to implement memory; for example, by connecting influencing elements in a delay buffer configuration such that previous values as well as current values can influence a target element [18], [26].

Spatial and temporal context is also important for choosing the appropriate parameters for a component model before aggregation, or when determining more abstract representations.

IV. CASE STUDY

In the following case study, we demonstrate the application of our methods to create and simulate a proof-of-concept hybrid element-based model of food security. First, we describe the high-level structure of an element-based model of food security. Then in Part I of the case study, we quantize rainfall data to provide time-series set values as input to the element-based model and to investigate effects of differing resolution of connected elements. In Part II of the case study, we integrate a quantized component model – a statistical emulator of crop yield – as a lookup table update rule in the element-based model, in combination with time-series set values quantized from the same input data used by the emulator. The results show the feasibility of integrating detailed component models and time-series model input via quantization in the element-based framework.

*A. Element-Based Model of Food Security*

The proof-of-concept element-based model for this case study includes an influence of precipitation on crop yield, where we demonstrate our methods of integrating time-series data and representations of component models to create the hybrid element-based model. Surrounding the component model are additional elements representing factors related to food security, shown by the influence diagram in **Fig. 4**. For all results shown in this case study, the hybrid element-based model is simulated according to a random-order sequential update scheme, where at each simulation time step one element is randomly chosen to be updated according to its update function [12]. The model is simulated repeatedly for a total of 1000 simulation runs, and the initial values are identical across simulation runs, representing one scenario. This number of simulation runs is sufficient to obtain a stable average of values across runs.

*B. Part I: Time-Series Set Values*

For the first part of this case study, we demonstrate quantizing a time series to set values of an input element in the element-based model. We test different variable value resolution for the time-series quantization and for downstream elements to analyze interactions between mixed-resolution elements. In this first part, we assume a simple relationship between precipitation and crop yield, where precipitation is the only positive influence on crop yield (**Fig. 4**, Part I), in order to focus on effects of quantization and mixed-resolution elements. We will build the complexity of this relationship in Part II.

We use weekly cumulative precipitation calculated from CHIRPS precipitation data for the Gambella region of Ethiopia for the time period from January 1, 2014 to April 30, 2018 [27]. This data is used to set values of the precipitation element over the course of the simulation. Each simulation time step corresponds to one week.

We quantize this precipitation data using either three discrete levels or nine discrete levels, where the quantization range is defined according to the minimum and maximum values of the precipitation data, and quantization thresholds are uniformly distributed across the value range. Values of the crop yield element are similarly represented with either three or nine discrete levels. Our reasoning for using three levels in the element-based model is to provide some representation of intermediate levels while minimizing computational complexity. However, using only three levels may not adequately represent changes in a value. Therefore, we compare quantization using three levels to quantization using nine levels, to investigate effects of a threefold difference in resolution on element interactions. We use odd numbers of discrete levels in order to have comparable levels representing 0%, 50%, and 100% at each resolution. Uniform quantization is chosen for simplicity in this proof-of-concept, but as previously discussed, future work could investigate non-uniform quantization based on context, qualities of the output, or available information.

Four simulation cases represent each combination of levels for precipitation and crop yield, shown in Table I, with simulation traces shown in **Fig. 5**. By simulating each combination of levels, we can investigate the effect of (1) equal resolution of the upstream (influencing) element relative to the downstream (target) element with Cases 1 and 3, (2) higher resolution of the upstream element relative to the downstream element with Cases 1 and 2, and (3) higher resolution of the downstream element relative to the upstream element with Cases 1 and 4.

TABLE I
SIMULATION CASES

|  | Precipitation Levels | Crop Yield Levels |
|---|---|---|
| Case 1 | 3 | 3 |
| Case 2 | 9 | 3 |
| Case 3 | 9 | 9 |
| Case 4 | 3 | 9 |

The initial value of the crop yield element is set to 50% in all cases: a value of 1 for three discrete levels, or a value of 4 for nine discrete levels. The precipitation element is an input in the element-based model and its values are set according to the quantized time-series data.

Simulation results show higher crop yield for Case 2 than Case 1, due to the higher frequency of zero values for three-level precipitation in Case 1.

Case 2 shows slightly higher peak crop yield than Case 3, due to differences in the distribution of next-state increments for crop yield calculated from the influencing precipitation value. In Case 2, the increment for crop yield is normalized to three levels, and the majority of increments were 1 across 1000 simulation runs; in Case 3, the increment for crop yield is normalized to nine levels, and the increments are more distributed because increments that would have normalized to 1 with three levels are distributed over intermediate values with nine levels. This result indicates that higher resolution elements show more nuanced changes, but overall trends are similar regardless of target element resolution given the same influencing element resolution.

Time-series set values derived from data will typically be of higher resolution than other elements in the model, but for completeness we include Case 4 to analyze the effects of lower-resolution upstream elements on higher-resolution downstream elements. The results are very similar to Case 1 (3 levels for both precipitation and crop yield); essentially, increasing resolution of downstream elements does not have much effect if upstream elements are lower resolution.

While this first part of the case study is sufficient for analyzing effects of resolution with the direct effects of precipitation on crop yield, this is not accurate to the actual relationship between precipitation and crop yield. Therefore, in Part II, we integrate a component model for crop yield that has precipitation as an input.

*C. Part II: Lookup Table Update Rule*

In this part of the case study, we use quantization to generate a lookup table update function representing a component model for crop yield.

We use a statistical emulator described in [6], [24] as the crop yield component model. Inputs to the emulator are primarily mean monthly precipitation and temperature for three summer months and annual $CO_2$ concentration. The emulator accounts for spatial grid cell effects and effects of soil type, and the output of the emulator is crop yield for a specified type of crop in metric tons per hectare. The tool described in [24] allows regional aggregation of the crop yield output.

The inputs to this crop yield emulator would not be expected to have upstream influences in the element-based model (i.e., other elements do not influence precipitation or temperature). Therefore, the values of crop yield produced by the component model could be quantized and used as time-series set values for the crop yield element in the element-based model, in which case the crop yield element is an element-based model input. However, a different component model may have upstream influences on its inputs when inserted into the element-based model, in which case the component model inputs must have corresponding elements in the element-based model. As a proof-of-concept for this methodology, we follow the strategy proposed earlier (**Fig. 3**) for mapping component model variables to the element-based model. The process specific to this case study is illustrated in **Fig. 6**.

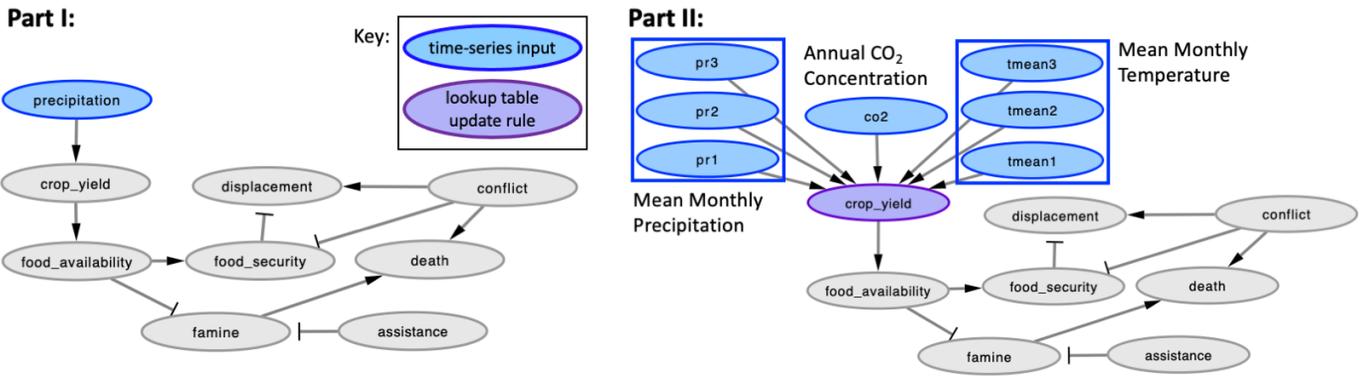

**Fig. 4.** Hybrid element-based models of food security.

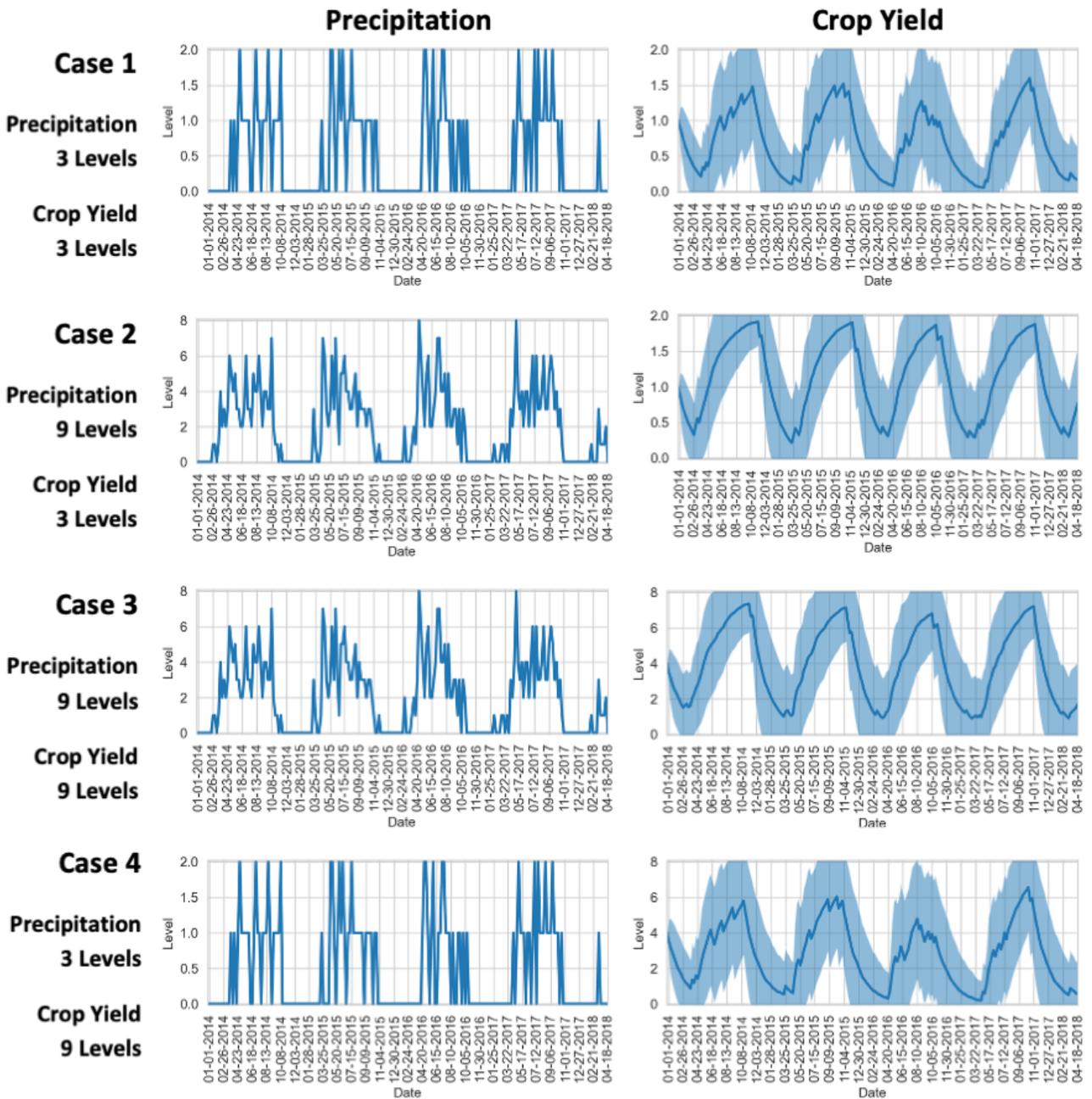

**Fig. 5.** Quantization of CHIRPS precipitation data for time-series input.

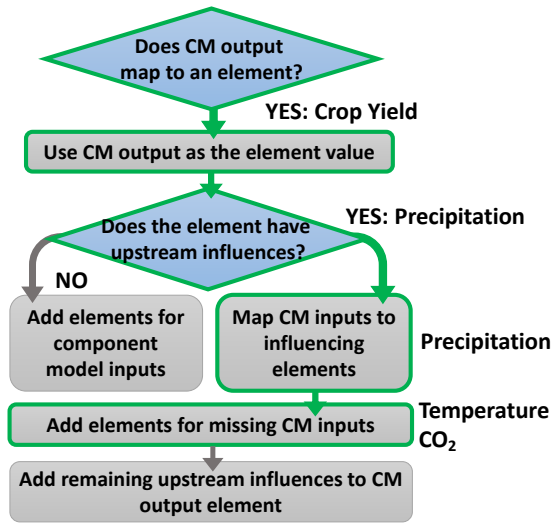

**Fig. 6.** Mapping of crop yield emulator variables to the element-based food security model.

Based on this process we map the crop yield emulator output to the existing crop yield element, map the emulator precipitation input to the existing precipitation element, and add temperature and $CO_2$ concentration as elements influencing crop yield, resulting in the hybrid element-based model shown in **Fig. 4**, Part II. Precipitation, temperature, and $CO_2$ concentration therefore act as inputs to the lookup table update function representing the crop yield component model.

While the emulator chosen for this case study provides aggregated output, the climate inputs are spatially disaggregated and combined with grid cell effects in the emulator functions. Because we cannot spatially disaggregate an element in the element-based model, we instead need to preprocess the component model inputs to the same spatial resolution as influencing elements. Therefore, we spatially aggregate climate inputs by taking the mean across a given region for each climate input from [24]: monthly temperature and precipitation, and annual $CO_2$ concentration. We then map these aggregated inputs to crop yield output from the [24] emulator for the Lund Potsdam-Jena managed Land (LPJmL) dynamic global vegetation and water balance model with the GFDL rcp8p5 climate scenario, aggregated at the country level. This effectively produces a spatially-aggregated regional emulator.

With input and output values at consistent spatial resolution, we use quantization to construct a lookup table update rule. We narrow the context of the component model to represent the relationship between climate inputs and maize yield for Ethiopia.

We define uniform quantization thresholds for each input and output based on a pre-determined number of levels and the range of values present in the data for each variable, as shown in Table II. Similar to Part I, we apply uniform quantization, and minimum and maximum quantization thresholds are defined based on the minimum and maximum values in the emulator data. Ranges for temperature and precipitation are defined as the minimum and maximum values across all three summer months in the climate data. Temperature, precipitation, and $CO_2$ concentration inputs are each quantized to three levels by defining two quantization thresholds uniformly spaced between the minimum and maximum values. Values greater than or equal to the minimum and less than or equal to the first threshold are quantized to a discrete level of 0; values greater than the first threshold and less than or equal to the second threshold are quantized to a discrete level of 1; values greater than the second threshold and less than or equal to the maximum are quantized to a discrete level of 2. While all inputs are quantized to three levels, we compare quantizing the maize yield output to three levels or nine levels, similarly defining the minimum and maximum values according to the range of values in the emulator output and defining uniformly spaced thresholds.

Then, we construct a table of all possible combinations of the discrete input levels – with three levels for each of the seven inputs, this results in 2187 input combinations. We fill in this table with the quantized maize yield output for input combinations present in the emulator data, merging the maize yield output for any duplicate input combinations by taking the median and rounding to the nearest integer level. While in this work we focus on effects of quantizing results already present in the emulator output, future work could explore methods of filling missing output values. Values could be obtained from the emulator, but there are associated disaggregation issues – the emulator does not take aggregated climate values as input, so we would need to either provide disaggregated climate inputs to the emulator equivalent to the missing aggregated input values, or determine a function representing the aggregated emulator to calculate the missing output values.

The complete lookup table is then used as the update rule for the element representing crop yield in the element-based model. We validate the lookup table update function in simulation by using the aggregated values for temperature, precipitation, and $CO_2$ from the emulator dataset (for the years 1971 – 2099) as time-series set values over the course of the simulation (**Fig. 7**), and comparing the simulated value of the crop yield element output to the original emulator maize yield output for the same time period (**Fig. 8**).

Due to the temporal resolution of the emulator data, each simulation step represents one year. Crop yield is initialized to 0 to match the initial value in the emulator output for the year 1971; all subsequent values are simulated according to the lookup table update rule. Simulation results are shown as the average across 1000 stochastic simulation runs with shaded area showing standard deviation across simulation runs.

In generating abstract representations of component models, we aim to use the fewest number of levels that still represents behavior. Using more levels for inputs results in a larger lookup table and requires more results from the component model. These validation results show that using the aggregate emulator as an update function in the element-based model recapitulates trends of the original emulator results, even with all variables in the lookup table quantized to 3 discrete levels.

TABLE II
QUANTIZATION THRESHOLDS

|  | Minimum | Intermediate Thresholds |  |  |  |  |  |  | Maximum |
|---|---|---|---|---|---|---|---|---|---|---|
| Temperature [°C] | 21.3 | 24.4 |  |  |  |  | 27.5 |  |  | 30.6 |
| Precipitation [mm/day] | 1.3 | 3.7 |  |  |  |  | 6.0 |  |  | 8.3 |
| $CO_2$ concentration [ppm] | 325.9 | 526.1 |  |  |  |  | 726.4 |  |  | 926.7 |
| Maize yield [t/ha] (3 levels) | 0.5 | 1.1 |  |  |  |  | 1.7 |  |  | 2.3 |
| Maize yield [t/ha] (9 levels) | 0.5 | 0.7 | 0.9 | 1.1 | 1.3 | 1.5 | 1.7 | 1.9 | 2.1 | 2.3 |

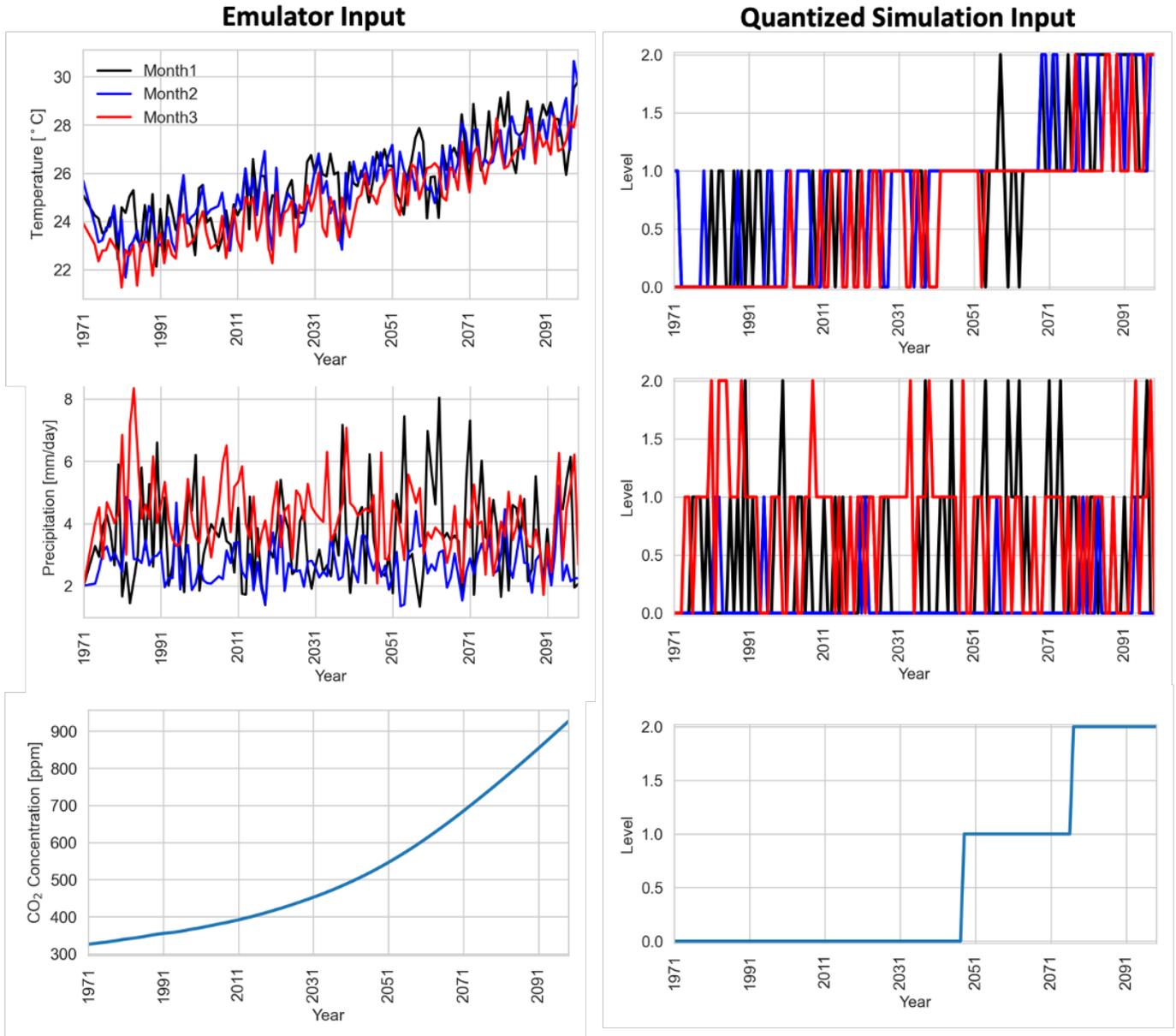

**Fig. 7.** Quantization of emulator inputs for element-based model input.

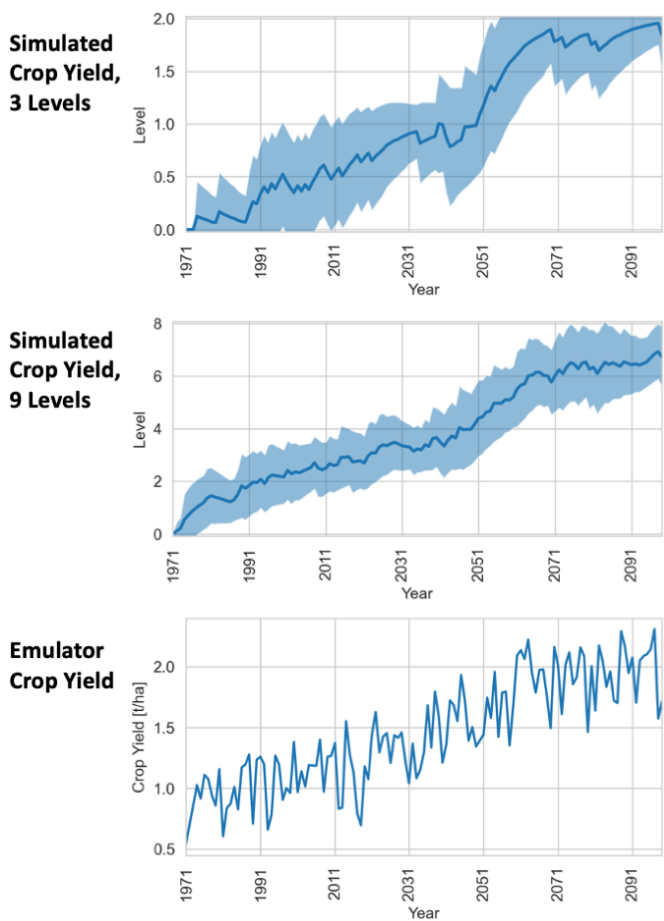

**Fig. 8.** Comparison of crop yield simulation results.

Note that the monthly values for temperature and precipitation were input as three separate elements each for temperature and precipitation, consistent with each simulation time step representing one year. Given an emulator that takes into account climate during any season (rather than just the three summer months), we could represent monthly temperature and precipitation with one element each, and have those elements influence crop yield to achieve higher than annual temporal resolution.

## V. Discussion

In building computational models, abstraction is necessary where detailed models are computationally impractical or where information is lacking. However, it is beneficial to utilize information from detailed models where they are available. Emulators are generally less computationally demanding than detailed process-based models, while still generating similar output, interpolating missing values, and potentially providing further abstraction by reducing the number of inputs or using approximated input values.

In this work, we presented a method of further abstracting emulators for integration as component models in a hybrid element-based framework with the following procedure: (1) mapping the emulator variables to elements in the element-based model, (2) spatially aggregating the emulator results, (3) converting the emulator input/output relationship into a discrete lookup table via quantization, and (4) applying the lookup table as an update rule in the element-based model for simulation of the complete system. As the hybrid element-based framework allows mixed-resolution elements in the same model, we also analyzed interactions between elements of different resolution, and compared emulator quantization with 3 or 9 discrete levels for the output. The results of our case study show that the quantized aggregated emulator simulated in the element-based framework recapitulates the results of the original emulator, even when quantized to 3 discrete levels for all variables. We therefore propose that this methodology enables simulation of systems with varying levels of information and availability of detailed models, through abstraction of component models and integration in a hybrid element-based framework.

As expected, some detail is lost when developing model abstractions. Blanc [5], [6] notes that emulators are better for long-term estimates because they capture variability over time with lower resolution. Additionally, in this work we spatially aggregated the results to a regional level, resulting in lower spatial resolution. While some aspects of detailed models are lost in the abstraction process, abstraction is often necessary to make complex system models more understandable and usable. There is value in abstraction even when information is available, not only to make analysis less computationally demanding, but also to aid interpretation of simulation results and to facilitate model exploration [9]. For example, statistical emulators trained on outputs from process-based models may lose mechanistic information in comparison to the process-based model, but can provide interpretability via model coefficients [5]. One advantage of a detailed mechanistic model is that effects of perturbations propagate through the system according to those mechanisms, and therefore a mechanistic model can be used to test various conditions and explore the response of the system. Detailed model analyses may reveal unexpected results through mechanistic interactions, while abstract analyses allow analysis of more and varied system components and their interactions, particularly when spanning multiple domains. Both approaches are valuable for exploring and understanding a system, and therefore using detailed models to inform abstract representations is a useful connection.

While in this paper we have presented a proof-of-concept of our methodology, further work is needed to address the complexities of integrating multiple component models, preserving component model metadata, missing component models, uncertainty metrics, and quantization methodology, as discussed below.

### 1) Multiple Component Models

There may exist multiple viable component models for a given concept, especially when working across levels of abstraction – for example, with the availability of maize, rice, wheat, and soybean emulators there is the question of how to combine these into a more abstract "crop yield" element representation [5], [6]. Models are often tailored for different applications, and clearly-defined context can aid in choosing the most appropriate model for a given purpose. But, even in the same context, there may exist multiple models that account for different factors or time scales in their estimates, causing differences in output [1], [5]. In some cases, it may make sense to aggregate results from multiple models and use differences in model output to estimate uncertainty [6]. We assumed deterministic update functions in this work, but future work could include probabilistic functions to represent multiple component models when context information is not sufficient to choose one component model.

  2) **Component Model Metadata**

When integrating component models, it is important to preserve component model metadata to enable integration of any updates to the component model or to inspect results of the original model if needed. Representations of model metadata are outside the scope of this work, but prior work has defined standardized representations for model metadata that could be applied to hybrid models that integrate component models [28], [29]. Additionally, with multiple component models in a hybrid element-based model, it is important to assume the same context for all component models and ensure consistency of any values shared among component models.

  3) **Missing Component Models**

Conversely to handling multiple models, there will be contexts where appropriate models do not exist, especially where data is less available. It may be possible to generalize models constructed for another context, but it is difficult to detect when a generalized model does not accurately represent behavior in a context for which it was not designed or tested. Ideally, an element-based model is informed by available knowledge, but in cases where information is not available, methods can be applied to infer unknown parameters or structure of the model given expected results for elements of interest [7], [30], [31] , [32], [33].

  4) **Uncertainty**

Calculating and preserving uncertainty metrics is a consideration as well, to quantify the information loss associated with quantization and relative to the original model. The most relevant metric for the quantization methodology presented in this work would be mean square quantization error (MSQE). It is expected that MSQE would decrease with higher value resolution, but as indicated in this work higher value resolution is not necessarily desirable or meaningful for element-based modeling. Future work could evaluate uncertainty metrics and their usefulness in discrete representations of component models.

  5) **Quantization**

While the methodology presented in this work is generalizable to other component models, there is a need for more automated quantization methods for integrating new component models or updating existing component models with new results. This is complicated by unique characteristics of each component model and the level of expertise required to define meaningful quantization thresholds [8]. As we showed in the case study, increasing resolution of target elements does not have much effect if the resolution of influencing elements is fixed; therefore, we can use the resolution of elements upstream of a component model to inform the resolution of component model inputs. Additionally, non-uniform quantization could be applied to better represent effects of component model input values on the output (e.g., thresholds could be set according to changes in the output value). However, non-uniform quantization must be accounted for in the effects of surrounding elements in the element-based model (i.e., when defining update functions). Further work on this method could also apply alternate techniques for merging output values for duplicate quantized input combinations or for filling missing values in the quantized lookup table.

## V. CONCLUSION

In this work we presented a methodology to integrate information from detailed component models and abstract relationships in a hybrid element-based model, enabling mixed-resolution modeling and simulation of systems with varying levels of information across component parts. In the context of modeling food security, we discussed and demonstrated the use of quantization and new simulation functionality in our element-based modeling framework to input precipitation data and to integrate a discrete-value representation of a statistical emulator of crop yield. While our proof-of-concept was a model of food security, the methodology is generally applicable to adapting other detailed models for simulation in the element-based framework. The results of this work therefore support the use of abstract representations of detailed models to enable computationally lightweight and interpretable modeling of systems spanning multiple domains.

ACKNOWLEDGMENT

This work is supported in part by DARPA award W911NF-18-1-0017.